# C++ Builder 6.0 CEDV code for files processing

*Yu.S.Tsyganov*,

*tyura@sungns.jinr.ru*

**Abstract**

A RAD C++ Borland's Builder 6.0 code **CEDV** under Windows XP has been designed for visualizing data obtained from heavy-ion-induced complete fusion reactions at the main U-400 cyclotron of FLNR. The main purpose of the code is processing the data from the experiments aimed at studying chemical properties of SHE. Data from the Dubna Gas Filled Recoil Separator could be processed too. Some subroutines for estimating statistical parameter are also presented; these are based on modified BSC (background signal combination) approaches.

**1. Introduction**

The experiments in low-energy physics are planned so that physically interesting effects mathematically are rare events, compared to others, whose probabilities are not small. Correspondingly, the purpose of the paper was the development of the formalism for the treatment of rare events, which definitely started for the recoil (ER) signal and after n-alpha particle signals is finished by the FF (fission fragment) signal in the same strip and vertical position of the focal plane PIPS detector (see. Eg.[1-3]). Classical models (eg. [4,5]) usually deal with the single imitator-signal rate to calculate (estimate) random events value $N_R$. Schematically it is shown in the Fig.1. It is evident that to a some approximation one could consider not only regular single background signals to built some combination, but also pairs, triplets etc. signals as primary backgrounds. Of course, it is assumed that the these second type background rate is much smaller than single signal rate and can not take into account analytically-only via direct processing of experimental data. There are some physical reasons for those kinds of signals. For instance, they are: all breakdown phenomena like Dee or even PIPS detector oxide rare breakdowns, as well as any RF pulse noises. Of course, in a real experiment e definite precautions should be taken, but sometimes their efficiency is not 100%.

In a sense, to some extent, background signals of this type may be considered as an intermediate stage from "thing-in-itself" (internal cause) to "thing-for-us" (obviously observed values).

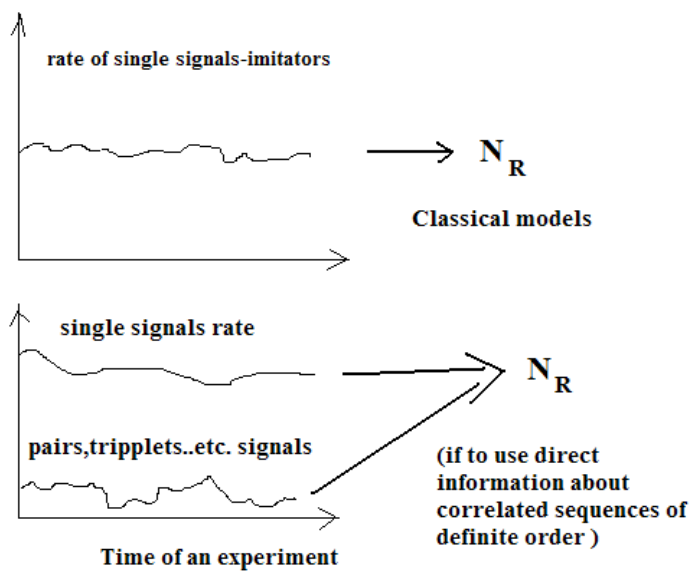

**Fig.1** Schematics of calculation of expectation number of random events in classical models and in the present approach

2. **Basic models to calculate the random coincidence probabilities and random event expectation.**

It was K.-H. Schmidt who first recognized the entire significance of probabilities estimates for the multi-chain events detected with a PIPS detector located at the focal plane of SHIP velocity filter recoil separator [4]. It developed a very frequently used by the experimentalist LDSC (linked decay signal combination) model. Another principal (basis) approach by V.B.Zlokazov [5] considers background signal combination (BSC) and test whether the signal sequence analyzed does fit in this concept or not.

In the same time, a trend of every experimentalist to use parameters which can be extracted directly from the given experiment and not only the rates of the single signals which are imitated the ER's, alpha particle and FF signals during the long term experiment. Moreover, when one apply "active correlations" method one can separate different signal groups according to the elapsed time [6,7]. Note, that according to the philosophy [4] there are some restrictions in application of the formulae from that paper. Namely, the background signal statistics should be not too extensive. That is, no any signals from the definite energy range should be present except for the signals included into the links. Generally speaking, it is preferable to apply the formulae from [6] in the case of the resulting NR is too small and much more less than 1. Otherwise, one can obtain more optimistic estimate, than realistic one. From the other hands, using the model [5] is quite questionable in the case of time interval from the end of alpha-group to the FF signal is much greater than the time duration from the recoil to finishing alpha particle signal. In part, this contradiction was solved in the modified model MBSC [5a].

In the case of signal combination α(n) –FF is under experimentalists interest, expectations to obtain together with this cluster starting α(n) imitator signal for a time $t_{α(n)-FF}$ and finishing FF signal for a time $t_{α(n)-FF}$ can be written as:

$$\overline{N}_R \approx N_{FF} \cdot P_{α(n)FF} \qquad (1).$$

Here, $P_{α(n)-FF}$ – is the probability to "detect" of group α(n) during a corresponding time interval before a FF signal for a time $t_{α-FF}$.

Having applying in the manner similar to [4, 5] Poisson like statistics for the mentioned signal parameters we obtain:

$$\overline{N}_R \approx N_{FF} \cdot (1 - e^{-\lambda_{α(n)} t_{α-FF}}) \qquad (2).$$

In (2) parameter $N_{FF}$ is the total number of FF signal for a given detector during the experiment.

### 3. File processor CEDV code for chemical and DGFRS experiments with actinide targets

For file processing of the data, obtained from the heavy ion reactions the **CEDV** (**C**hemistry **E**xperiments **D**ata **V**isualization) RAD C++ Builder under Windows XP code has been designed. It allows to process files from the experiments performed at FLNR (JINR) in the field of superheavy elements [8]. One of the output results of its application is a list of α-α-α-FF correlations under definite time conditions. In the Fig.1 **3D** example is shown for three alpha particle signal for a given detector[1].

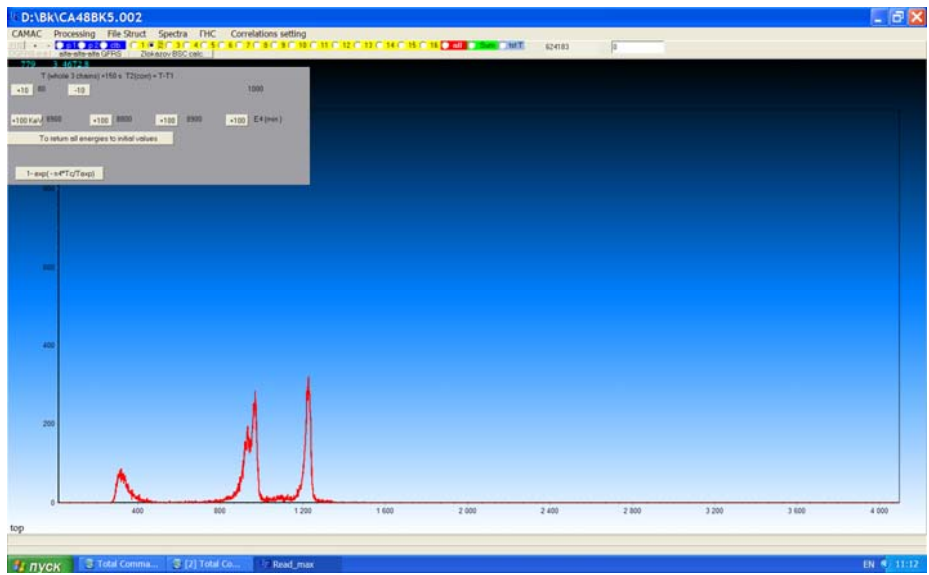

**Fig. 2a** Spectrum obtained from the experimental file. Left-upper corner - parameters for α-α-α and α-α-α-α correlations search (for a given detector) is shown. (for 16 pairs(top + bottom) with close to 4π geometry- 16 buttons in the upper panel) Bottom button at the panel- is one to calculate $N_R$ parameter after correlation search process is finished.

---

[1] Whereas, any **2D** alpha-alpha picture has negligible statistical significance! (Fig.2j)

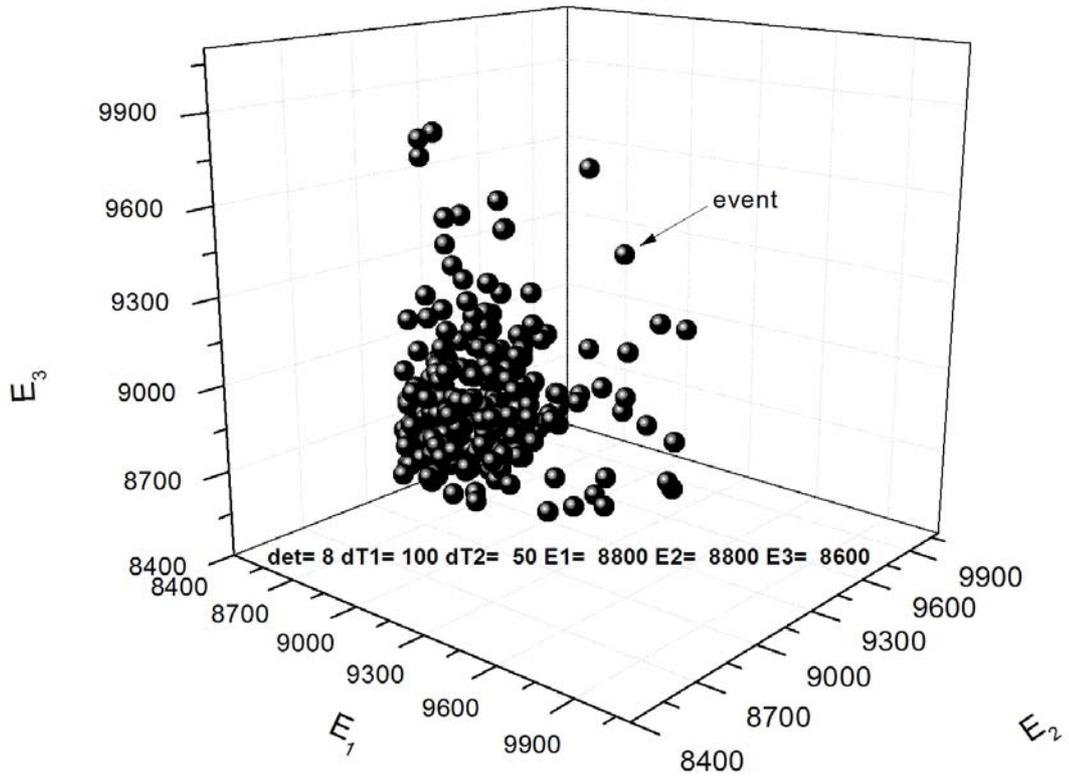

**Fig.2 b** 3D picture of α-α-α group[2]. Detected event indicated by the arrow. Left part-background signals for E1=E3=8800 KeV and E2=8600 KeV.

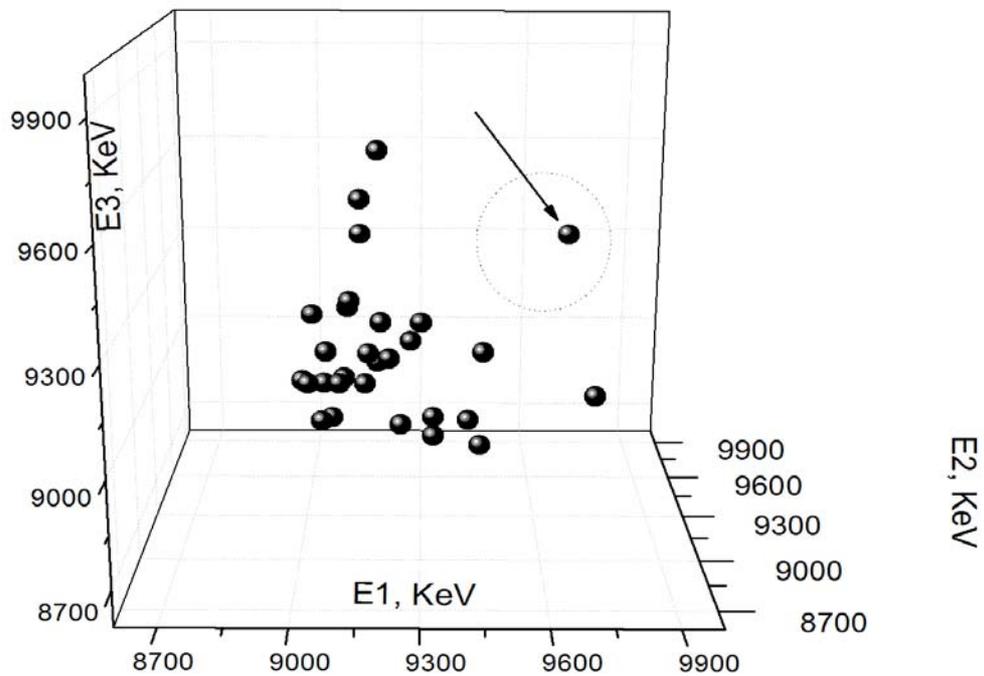

**Fig.2c** *3D picture of α-α-α group. Detected event indicated by the arrow. Left part-background signals.*
*For chain:* ***9644 KeV → 8935 → 9549 → 8817*** *987671.312 64.812 6.562 815.625)*

---

[2] An interactive mode. In principle, *on Timer* mode is possible easy for any on-line experiments

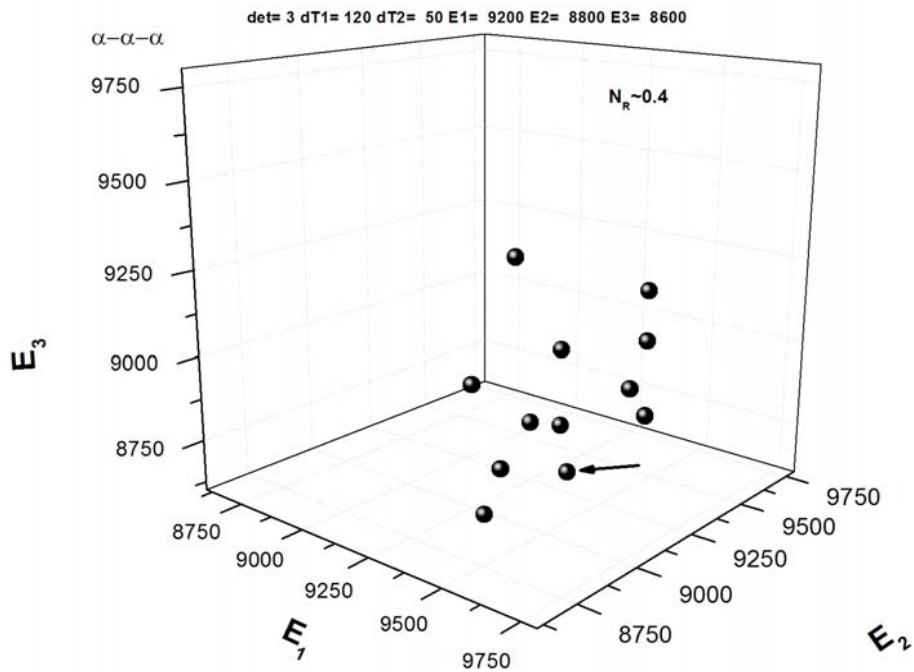

**Fig.2d** ..the same for another detector..( More worthier situation with $N_R$). Only three alpha-particles are taken into account

If one use formula (2) for the given case the value of expectation of $N_R \approx 0.147$ for incoming parameters: 9300, 8800 and 9300 KeV as low energy limits and time intervals $\leq 100$ s for first-second alpha particle interval and $\leq 50$ s for second-third interval, respectively. In the Fig.2 low limit for $E_1=E_3$ is considered as a parameter under condition of fixed parameter $E_2=8800$ KeV. For the sake of comparison, one can see in the Fig.3 the analogues picture from the DGFRS experiment, when the "active correlations" method to suppress the background signals is applied. Of course, vertical position signal is also taken into account.

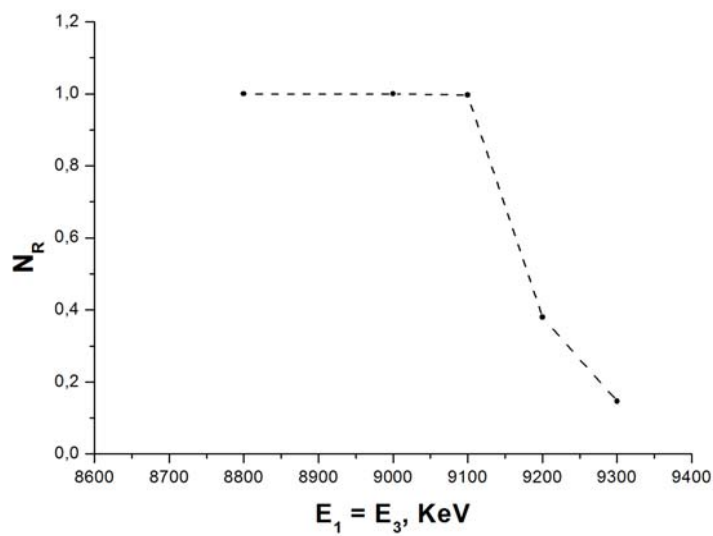

**Fig.2e** Dependence of $N_R$ parameter against the lower limits $E_1$ and $E_3$ for the fixed value $E_2=8800$ KeV.

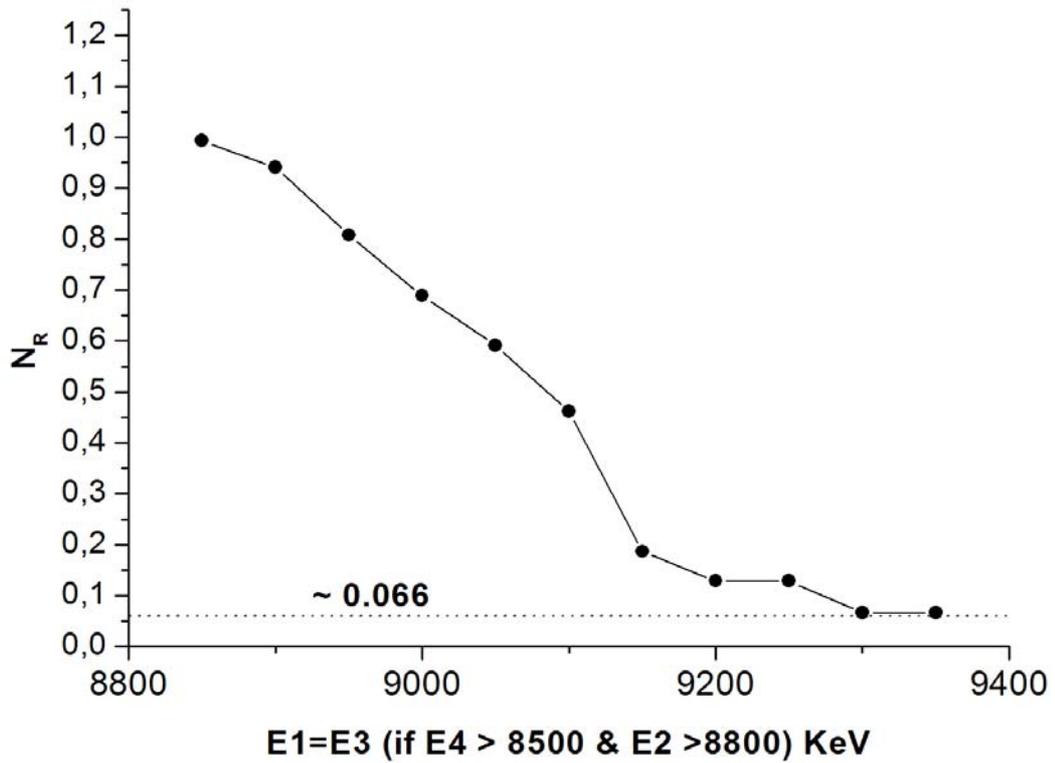

**Fig.2f** The same as 2e, but under additional condition $E_4 > 8500$ KeV. Asymptotic level of ~0.066 is shown by dotted line. (det. #8[3])

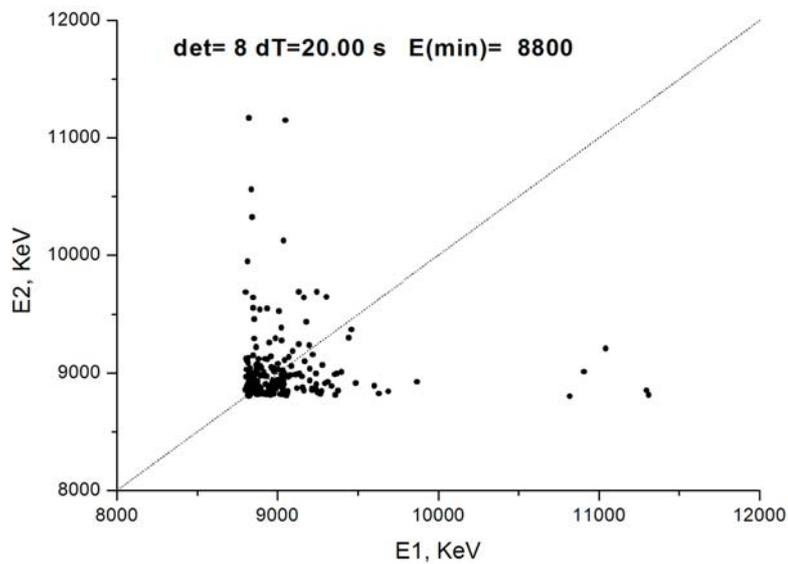

**Fig.2j** 2D picture for alpha-alpha correlated chains within 20 s time interval

---

[3] Having considering neighbor left-right detectors and taking into account ~ 85% efficiency per pair and 92% full efficiency relatively $4\pi$ [8], $N_R$ value will be slightly higher. Namely, it is equal approximately to 1.16. In means that $N_R \approx 0.066 \cdot 1.16 = 0.077$.

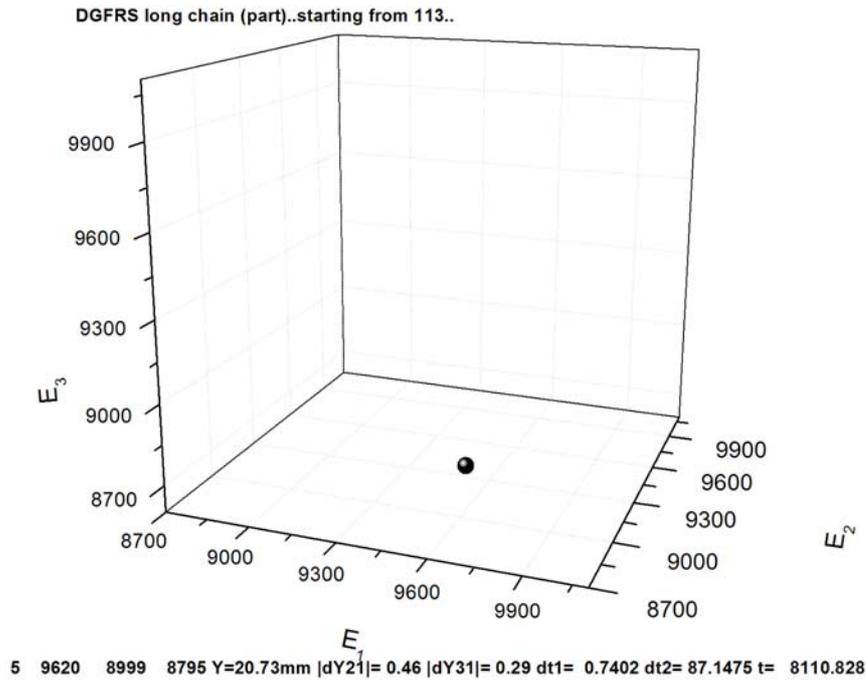

**Fig.3** One decay event from the DGFRS experiment [9] - demonstration of background-free conditions due to application of "active correlations" method!

## 4. On the issue of using the edge stability of statistical estimates

There is reasonable to consider some additional statistical criteria together with offently used $N_R$ parameter. It seems, to this aim consideration not only the mentioned parameter, but also first derivatives according to two basic variables, namely – time and energy are useful. …
For instance, in the Fig.4 the dependence $\frac{\partial N_R}{\partial E}$ against $N_R$ is shown. Rapid growth of the parameter in the right zero vicinity, probably, manifests the statistical parameter estimate non-stability region. To a first approximation it is proposed that simple relation $|\frac{\partial N_R}{\partial E} \cdot \Delta E| << N_R(E)$ and $|\frac{\partial N_R}{\partial t} \cdot \Delta t| << N_R(t)$ as a required condition to speak about parameter $N_R$ as a stable estimation. At the same first step one can also considered $\Delta E$ interval as one related with FWHM value of a detector and left edge energy as a critical point. Choice of $\Delta t$ value, probable, should obey more branched scenario.

Another test (sufficient condition) for validity of statistical significance estimate was proposed in [10]. In fact, some re-norm factor is introduced. It relates with the given configuration of the detected event and the whole efficiency to detect a single alpha particle. Stable estimate was defined as one corresponds to the case of the parameter of $N_R/P_{Conf} <<1$. Here $P_{Conf}$ is a probability of a definite configuration of an event.

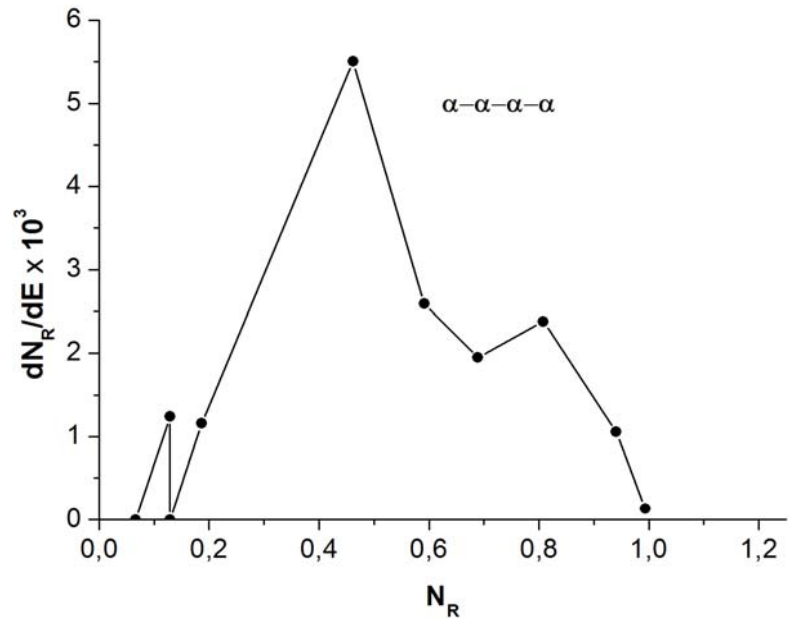

**Fig.4** Phase curve for α-α- α-α chain.

5. **Summary.**

Approach to calculating random expectation value for heavy ion induced complete fusion reactions has been proposed. It is applicable if a number $N_{α(n)}$ of α(n) combination signals can be directly extracted from the experimental data for the energy and time intervals of interest. Of course, it is assumed that $N_{α(n)} \gg 1$, as an ideal case. And, of course, there are some uncertainties, when time interval between last alpha particle signal and FF signal is compared with that from first to finishing alpha particle signal. But this is not the problem when only an order of magnitude estimate is of interest for experimentalist. In this case an effective time can be introduced into consideration by any way to overcome the mentioned problem. One can conclude that the chain shown in the Fig.2c is statistically significant, although with the remark, that the presented estimate may be unstable to some extent. In this case it essential whether the experimentalist considers the measured event as being known before the experiment or not. In the fist case, the statement about statistical significance of the event will follow without doubt. In contrast, no definite conclusion can be drawn for the event shown in Fig.2d. Additional thorough analysis is required, probably not just with statistical approach. As concerning the stability of the statistical estimates, it is the first time this point is introduced. Of course, the author welcomes anybody filling this definition with a new sense.

Since the major drawback of the described statistical estimate approach is an extension to a low level of background signals combinations, one should obtain not only the mean value of $N_R$, but also left and right confidence intervals for that estimate.

Additionally, three general conclusions can be drawn here, namely:

1. The event from $^{243}$Am+$^{48}$Ca reaction one can consider as a true event (Fig.6; see below);
2. The second event from $^{249}$Bk+$^{48}$Ca reaction should be considered as background nature;
3. Under definite circumstances, especially with the additional positive analysis, non-related with a statistical one, the first event (Fig. 2c) can be considered as a candidate to the event.

The author is indebted to Drs. A.V.Yeremin, M.L.Chelnokov, A.V.Isayev and Prof. S.N.Dmitriev for their assistance and fruitful discussions on the item reported in this paper.

**Supplement 1. Using VMRIA code to estimate peak content**

To estimate content of any nuclide in the peak area, ***VMRIA*** code has been designed and described in the Ref. [11]. To apply it in our case, ***CEDV*** code generates text-files in inter – active mode, acceptable by ***VMRIA***. In the Fig.5 an example of application is shown for heavy-ion induced nuclear reaction $^{243}$Am+$^{48}$Ca→115*(see ref. [8] too). Authors (present paper &[11]) plane to improve VMRIA code to satisfy more exactly the experimental conditions of the chemical experiments.

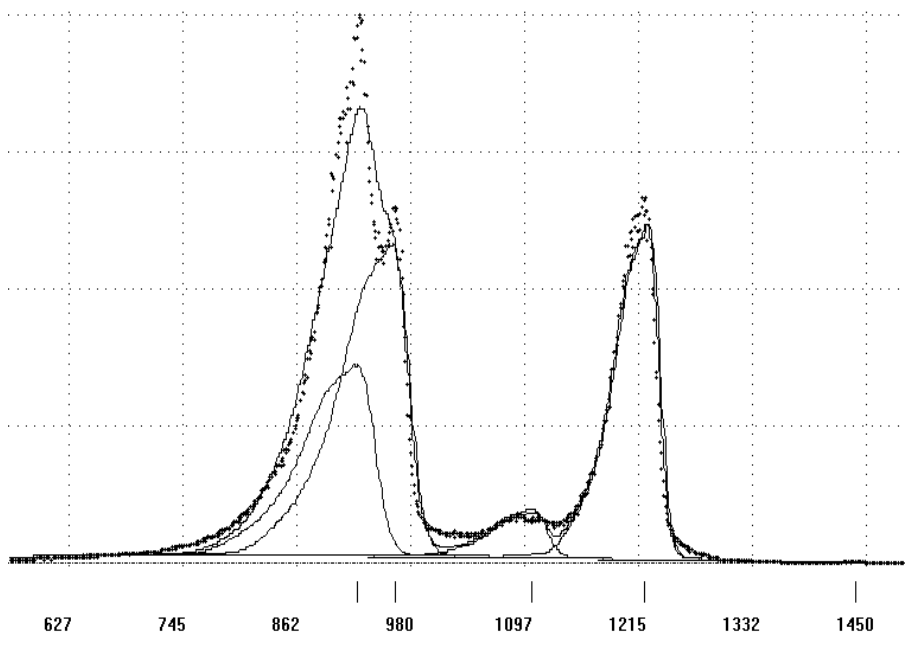

**Fig.5** Measured spectrum (dots) and different peaks (line) according to VMRIA processing from $^{243}$Am+$^{48}$Ca reaction

**Supplement 2. Search for short ER-α (or/and α-α) chains for DGFRS files.**

Sometimes a quick search of presence or absence of short correlated sequence is strongly required in the given file or group of files. To this aim in the CEDV code an appropriate menu button is designed for the DGFRS experiment file format. On pressing this menu item, short correlations like alpha-alpha can be extracted according to energy-time-position information.

Very useful information corresponds the case of ER-alpha chain for $^{217}$Th nuclide obtained in the $^{nat}$Yt+$^{48}$Ca[4] nuclear reaction (Fig.'s 7a,b), due to namely this reaction and nuclide is used for calibration purposes. Of course, it is assumed that all calibration parameters are ready for use. Below the figure a part of the resulting text buffer and meaning of the variables are presented. Of course, it is assumed that all calibration parameters are ready for application at that moment.

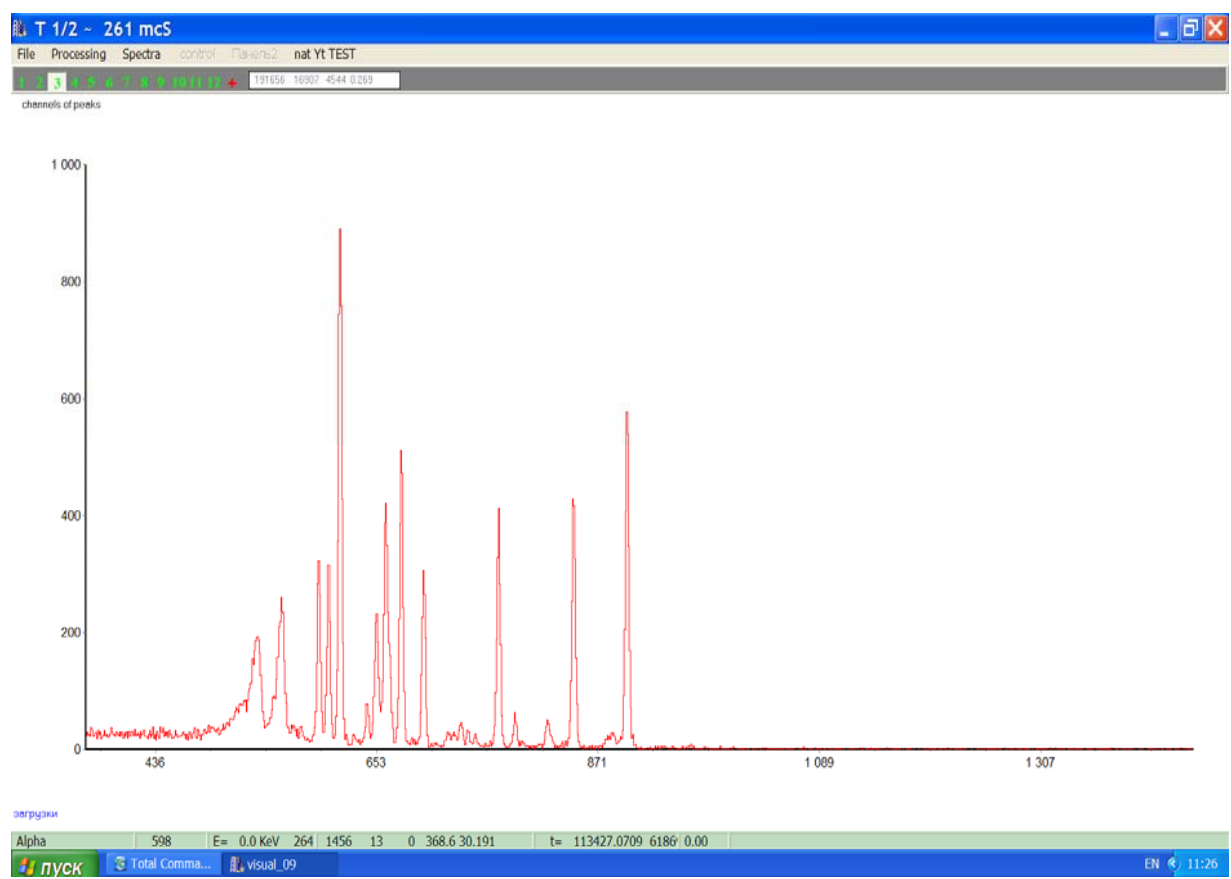

**Fig.7a** Spectrum from $^{nat}$Yt+$^{48}$Ca→Th* reaction. Peak under interest is in he right hand position in the spectrum ($^{217}$Th isotope). Rough estimate value of T1/2 is in the Code form caption (left-upper).

---

[4] One from the DGFRS calibration complete fusion nuclear reactions

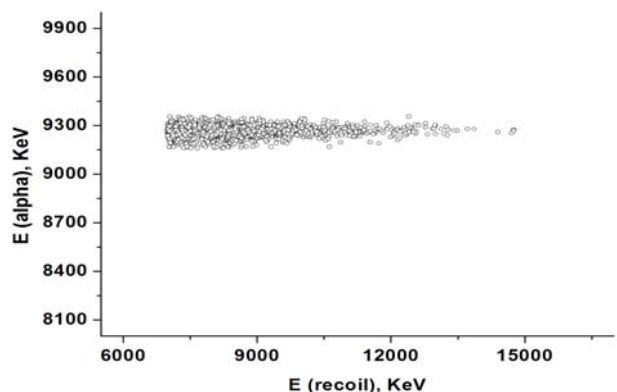

**Fig.7b** Correlation picture for extracted ER-alpha chains

```
4  485  8338  9320   0   91207.640 dt= 399
8  491  7252  9279   0   91219.555 dt= 899
1  505  8725  9242  -3   91227.204 dt= 500
4  491  7670  9289   0   91231.470 dt= 399
1  491  7369  9263  -4   91235.651 dt= 299
2  480  9735  9282  -2   91238.907 dt= 1499
```

Here are:
1st column - strip number;
2nd column – TOF amplitude of ER in channels;
3rd ...        - ER energy signal amplitude;
4th            - alpha particle energy signal amplitude;
5th            - deviation in vertical position in pixels ( one pixel= 40/130 mm);
6th            - elapsed time;
7th            - time difference ER-alpha.

## Supplement 3. Reaction Am+ $^{48}$Ca->115…->111…

The same (Fig.2) picture was created by ***CEDV*** for Am+$^{48}$Ca . It goes without saying, that background group is outside the event location area. Fig.6)

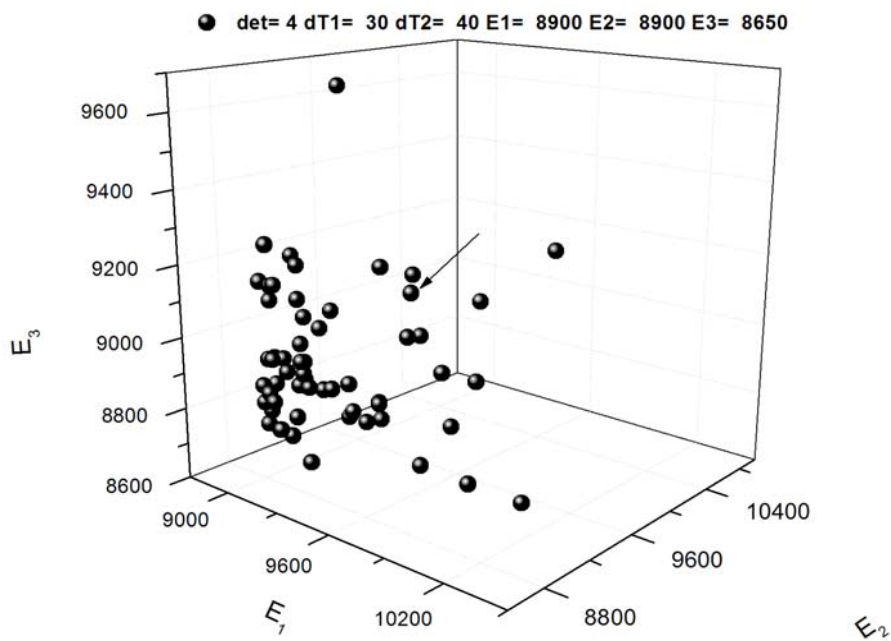

a)

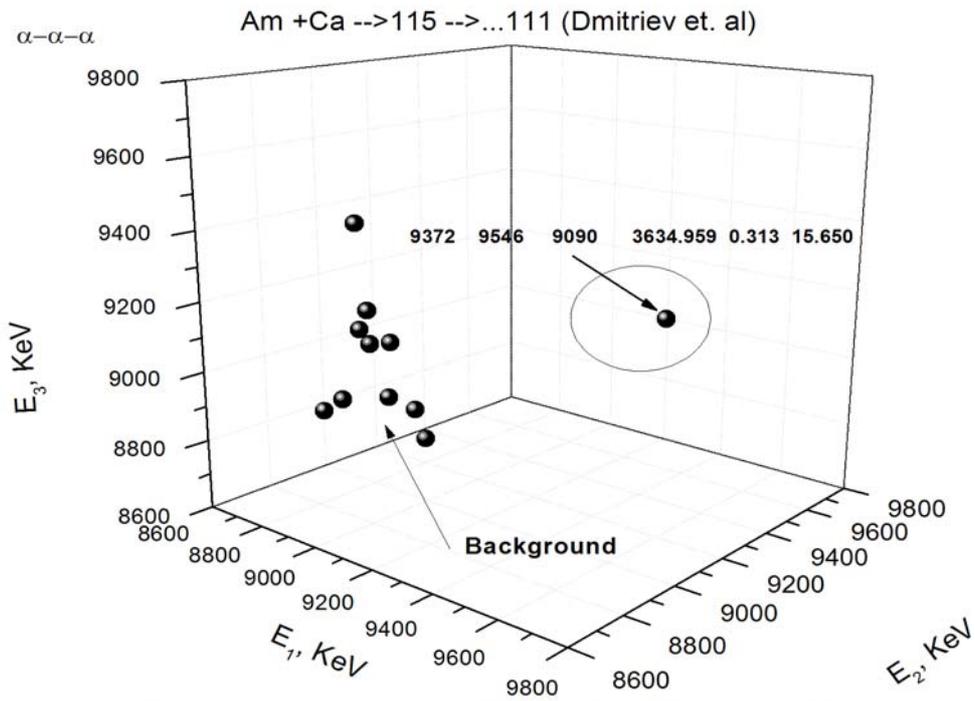

b)

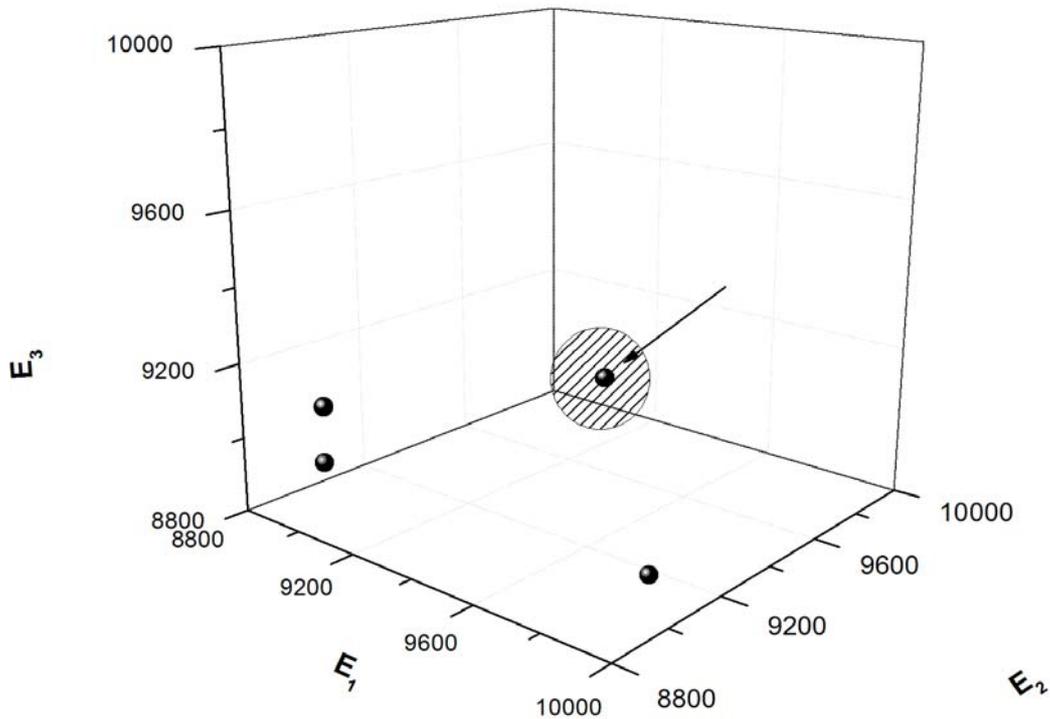

c)

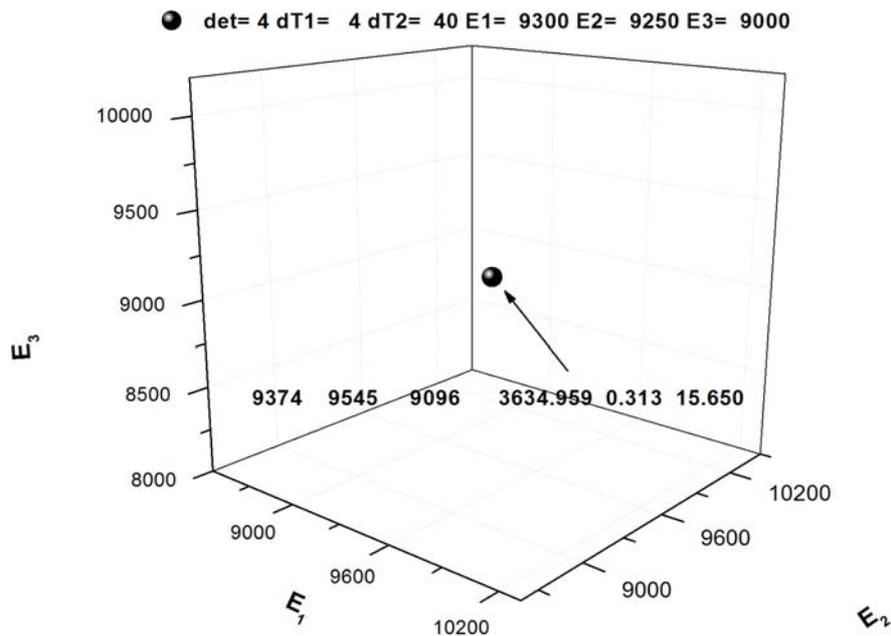

**Fig.6** *3D* correlation pictures for $^{243}$Am+$^{48}$Ca reaction [8]. The event is shown by arrow.